\documentclass[11pt]{article}
\input epsf

\def \in{\leftskip = 40 pt\rightskip = 40pt}

\def \out{\leftskip = 0 pt\rightskip = 0pt}

\def\NPB{{\em Nucl. Phys.} B}
\def\PLB{{\em Phys. Lett.}  B}

\def\PRD{{\em Phys. Rev.} D}

\def\be{\begin{equation}}
\def\ee{\end{equation}}
\def\bea{\begin{eqnarray}}
\def\eea{\end{eqnarray}}

\def\mbar{{\overline{m}}}

\def\frak#1#2{{\textstyle{{#1}\over{#2}}}}

\def\btil{\tilde b}  
  
\def\dtil{\tilde d}  
\def\etil{\tilde e}   
\def\gtil{\tilde g}

\def\ttil{\tilde t}
\def\util{\tilde u}

\def\tautil{\tilde \tau}

\def\nutil{\tilde \nu}

\def\MeV{\hbox{MeV}}
\def\GeV{\hbox{GeV}}
\def\TeV{\hbox{TeV}}
\def\half{\frac{1}{2}}
\def\lf{16\pi^2}

\def\vev#1{\mathopen\langle #1\mathclose\rangle }
\def\nn{\nonumber\\}
\def\DRED{\ifmmode{{\rm DRED}} \else{{DRED}} \fi}
\def\DREDD{\ifmmode{{\rm DRED}'} \else{${\rm DRED}'$} \fi}
\def\NSVZ{\ifmmode{{\rm NSVZ}} \else{{NSVZ}} \fi}

\def\pa{\partial}

\def\sic{supersymmetric}

\def\sy{supersymmetry}
\def\sic{supersymmetric}

\def\phib{\overline{\phi}}

\def\semi{;\hfil\break}

\begin{document}
\begin{titlepage}
\begin{flushright}
LTH 655\\
hep-ph/0507193\\
\end{flushright}

\vspace*{3mm}

\begin{center}
{\Huge
Anomaly Mediation, Fayet-Iliopoulos $D$-terms 
and precision sparticle spectra}\\[12mm]

{\bf R.~Hodgson, I.~Jack, D.R.T.~Jones}\\

\vspace{5mm} 
Dept. of Mathematical Sciences,
University of Liverpool, Liverpool L69 3BX, UK\\
\vspace{8mm}
{\bf and G.G.~Ross}

\vspace{5mm}
The Rudolf Peierls Centre for 
Theoretical Physics,
Oxford University, 1 Keble Road, Oxford OX1 3NP, UK\\
\end{center}
\vspace{3mm}

\begin{abstract}
We consider the sparticle spectra that arise when 
anomaly mediation is the source of \sy-breaking and 
the tachyonic slepton problem is solved by a Fayet-Iliopoulos $D$-term. 
We also show how this can lead to a minimal viable extension of anomaly 
mediation, in which the gauge symmetry associated with this 
$D$-term is  broken at very high energies, leaving as 
its footprint in the low energy theory only the required $D$-terms and 
seesaw neutrino masses. 
\end{abstract}

\vfill

\end{titlepage}

\section{Introduction}

Increasing precision of sparticle spectrum calculations is an important
part of theoretical preparation for the LHC and the ILC. Much of this
work has concentrated on the MSUGRA   scenario, where it is
assumed that the unification of gauge couplings at high energies  is
accompanied by a corresponding unification in both the soft
\sy-breaking  scalar masses and the gaugino masses; and also that the
cubic scalar interactions  are of the same form as  the Yukawa couplings
and related to them by a common constant  of proportionality, the
$A$-parameter. This paradigm is not, however, founded on  a compelling
underlying theory and therefore it is worthwhile exploring other 
possibilities.

In this paper we focus on Anomaly Mediation (AM)~\cite{lrrs}-\cite{Ibe:2004gh}. 
This is a framework in 
which  a single mass parameter determines the $\phi^*\phi$, $\phi^3$ and
$\lambda\lambda$ \sy-breaking terms  in terms of calculable and
moreover renormalisation group (RG) invariant functions of the 
dimensionless couplings, in an elegant and predictive way; too
predictive, in fact,  in that the theory in its simplest form leads to
tachyonic sleptons and  fails to accommodate the usual
electroweak vacuum state. There is a natural  solution to this, however,
which restores the correct vacuum while retaining  the RG invariance
(and hence the ultra-violet insensitivity) of the  predictions. This is
achieved simply (and without introducing another source of explicit 
\sy-breaking) by the introduction of a Fayet-Iliopoulos (FI) $D$-term  or
terms. 

This possibility was first explored in
detail in  Ref.~\cite{jja}, and subsequently by a number of 
authors~\cite{Arkani-Hamed:2000xj}-\cite{Ibe:2004gh}. The
main purpose of this paper  is to present the most precise spectrum
calculations to date in the  AMSB scenario. We also show how the low
energy theory employed  can arise in a natural way from a theory with an
additional anomaly-free $U_1$ broken  at a high scale. 
We examine the decoupling in this case and show how only the soft mass 
contributions from the $D$-terms remain, which can naturally 
eliminate the tachyonic slepton problem. This provides a minimal extension of 
anomaly mediation.

In the original scenario of Ref.~\cite{jja}, FI  terms corresponding to two
distinct $U_1$ groups were  introduced, one being the standard model
$U_1$, and the other  the second mixed-anomaly-free (or completely
anomaly-free if right-handed neutrinos are included) $U_1$ admitted by
the  MSSM. This $U_1$ may be chosen to be $B-L$~\cite{Arkani-Hamed:2000xj}, 
or some linear combination of it and  
the MSSM $U_1$~\cite{jja}. Or, as emphasised in
Ref.~\cite{mwells}, a single new $U_1$ may be employed if the charges are 
chosen appropriately. If these FI terms are added to the masses 
with constant coefficients 
(as in Eq.~\ref{eq:AE} below) rather than as genuine gauge linear  $D$-terms, 
then as discussed  in Ref.~\cite{jja} and at more length in 
Ref~\cite{mwells}, the choices made in Refs~\cite{jja}-\cite{mwells} are 
simply reparametrisations of each other. As indicated above, we will see how 
this scenario can emerge naturally at low energies in a specific theory 
with an additional gauged  anomaly-free $U_1$.

\section{The Minimal Supersymmetric Standard
Model}

The MSSM is defined by the superpotential:
\be
W =  H_2 Q Y_t t^c + H_1 Q Y_b b^c  + 
H_1 L Y_{\tau} \tau^c + \mu H_1 H_2 
\ee
with soft breaking terms:
\bea
L_{\rm SOFT} & = & \sum_{\phi}
m_{\phi}^2\phi^*\phi + \left[m_3^2 H_1
H_2 + \sum_{i=1}^3\half M_i\lambda_i\lambda_i  + {\rm h.c. }\right]\nn
& + & \left[H_2 Q h_t  t^c  +
H_1 Q h_b b^c  + H_1 L h_{\tau}\tau^c
+ {\rm h.c. }\right]
\eea
where in general $Y_{t,b,\tau}$ and $h_{t,b,\tau}$ are 
$3\times 3$ matrices. We work throughout in the approximation that the Yukawa 
matrices are diagonal, and neglect the Yukawa couplings of the 
first two generations.

\section{The AMSB Solution}

Remarkably the following results are RG invariant:

\bea
M_i & = & m_0 \beta_{g_i}/{g_i}\nn
h_{t,b,\tau} & = & -m_0\beta_{Y_{t,b,\tau}}\nn
(m^2)^i{}_j & = & \frac{1}{2}m_0^2\mu\frac{d}{d\mu}\gamma^i{}_j\nn
m_3^2 & = & \kappa m_0 \mu - m_0 \beta_{\mu}
\label{eq:AD}
\eea
Here $ \beta_{g_i}$ are the  gauge $\beta$-functions, $\gamma$ the  
chiral supermultiplet anomalous dimension, and $\beta_{Y_{t,b,\tau}}$ 
are the  Yukawa 
$\beta$-functions. Moreover, the RG invariance is preserved if we replace 
$(m^2)^i{}_j$ in Eq.~\ref{eq:AD} by 
\be 
(\mbar^2)^i{}_j =  \frac{1}{2}m_0^2\mu\frac{d}{d\mu}\gamma^i{}_j,
+kY_i\delta^i{}_j,
\label{eq:AE}
\ee
where $k$ is a constant and $Y_i$ are charges corresponding 
to a $U_1$ symmetry of the theory with no mixed anomalies 
with the gauge group. Of course the $kY$ term corresponds in form to 
a FI $D$-term.       
The expressions 
for $M$, $h$ and $m^2$ given in  Eq.~\ref{eq:AD} 
are  obtained if the only source
of breaking is a vev in the supergravity
multiplet itself: the AMSB scenario 
($m_0$ is then the gravitino mass).  Note the  
parameter $\kappa$ in the solution for $m_3^2$; some treatments 
in the literature omit this term (based on top-down considerations). 
However, Eq.~\ref{eq:AD} is RG invariant for arbitrary $\kappa$ 
and so we retain it. This means that $m_3^2$ will be determined in the 
usual way by the electroweak minimisation. In the following two sections we 
will show how Eq.~\ref{eq:AE} can arise via spontaneous breaking 
of a $U'_1$ symmetry.

\section{Anomaly-free $U_1$ symmetries}
The MSSM (including right-handed neutrinos) admits 
two independent generation-blind anomaly-free $U_1$ symmetries. 
The possible charge assignments are shown in Table~1. 
\begin{table}
\begin{center}
\begin{tabular}{|c c c c c c|} \hline
$Q$ & $u^c$ & $d^c $ 
& $H_1$ & $H_2$ & $\nu^c$ \\ \hline
& & & & & \\ 
$-\frac{1}{3}L$ & $-e-\frac{2}{3}L$  & $e+\frac{4}{3}L$ 
& $-e-L$ & $e+L$ & $-2L-e$ \\ 
& & & & & \\ \hline
\end{tabular}
\caption{\label{anomfree}Anomaly free $U_1$ symmetry for arbitrary 
lepton doublet and singlet charges $L$ and $e$ respectively.}
\end{center}
\end{table}

Note that an anomaly-free, flavour-blind $U_1$ necessarily 
corresponds to equal and opposite charges for $H_{1,2}$ and hence 
an allowed Higgs $\mu$-term; for an attempt at a 
Froggatt-Nielsen style origin for the Higgs $\mu$-term
using a {\it flavour dependent\/} $U_1$, see for example 
Ref.~\cite{jjw}. One of the attractive features of 
anomaly mediation is that squark/slepton mediated flavour changing neutral 
currents are naturally small; this feature is preserved 
by a  generation-blind $U_1$ but not by a flavour dependent $U_1$, so 
we stick to the former here. 

The SM gauged  $U^{SM}_1$  is  $L=1, e = -2$; this $U_1$ is of 
course anomaly free even 
in the absence of $\nu^c$.  $U_1^{B-L}$  is 
 $L = -e = 1$; in the absence of $\nu^c$ this would have $U_1^3$ and
$U_1$-gravitational  anomalies, but no mixed anomalies with the SM gauge
group, which would suffice to maintain  the RG invariance of the AMSB
solutions.  We can introduce  FI  terms for both $U^{SM}_1$ and 
$U_1^{B-L}$,   or for $U^{SM}_1$ and a linear   
combination of them\cite{jja}, or indeed simply have a single  $U'_1$ 
with the same sign for $L$  and  $e$~\cite{mwells}. We will follow
Ref.~\cite{mwells} here; however  in the decoupling scenario described
in the next section, the low energy theories  corresponding to the
single $U'_1$ case and and the double   $U_1$ case
of Ref.~\cite{jja} are simply reparametrisations of each
other. 

\section{Spontaneously broken  $U'_1$}\label{sec:amsb}

With the MSSM augmented by an additional $U'_1$, it is natural 
to ask at what scale this $U'_1$ is broken. It is possible that this scale 
is at around $1\TeV$~\cite{Erler:1999nx}; here, however, we concentrate on the 
idea that it is broken at very high energies and that the only 
low energy remnant of it is the set of FI-type terms that we require.

It would be natural to think that if a $U'_1$ is broken at some high
scale $M$ then,  by the decoupling theorem, all effects of the $U'_1$
would be suppressed at energies  $E << M$ by powers of $1/M$. We shall
see that with a FI term this is not the case  and it is quite natural
for there to be $O(M_{\hbox{SUSY}})$ scalar mass contributions arising 
from the presence of the FI term. 

It is straightforward to construct a  model with an  
additional gauged  $U'_1$  in such a way that the only effect
on the low-energy theory is the appearance of the  FI  terms we require.

We introduce a pair of MSSM singlet fields  $\phi, \phib$  with $U'_1$ charges 
 $q_{\phi, \phib} = \pm (4L+2e)$ and a gauge singlet  $s$, 
with a superpotential
 \be
W = \lambda_1 \phi\phib s + \half\lambda_2 \nu^c \nu^c \phi.
\ee 
The choice of charges is essentially determined by the requirement that the 
$\phi,\phib$ fields decouple 
from the MSSM while  generating a large mass for $\nu^c$. 

The scalar potential takes the form:
\bea
V &=& 
m_{\phi}^2\phi^*\phi + m_{\phib}^2\phib^*\phib+ \cdots \nn
&+& \frac{1}{2}\left[\xi - q_{\phi}(\phi^*\phi- \phib^*\phib)
- \sum_{\hbox{matter}}e_i \chi^*_i\chi_i\right]^2 + \cdots..
\label{eq:sp}
\eea 
where $\chi_i$ stands for all the MSSM scalars, and $e_i$ their $U'_1$ charges, 
and we have introduced a  FI  term for  $U'_1$. We will 
take $\xi > 0$, $q_{\phi} > 0$ and 
$\xi >> m_0^2$, and assume  that the scalar masses in 
Eq.~\ref{eq:sp} (apart from the Higgs $\mu$-term) and other 
\sy-breaking terms in the theory are the anomaly-mediation contributions. 
We now proceed to minimise the scalar potential. As we shall see, this will 
result in a vev for $\phi$ of order $\sqrt{\xi}$; this means that 
the appropriate scale at which we should minimise the potential 
is also around $\sqrt{\xi}$. As a consequence, 
we of course include at this stage the $U_1'$ contributions in the 
anomalous dimensions of the fields.
It therefore follows that as long as $\lambda_{1,2}$  are
somewhat smaller than  the $U'_1$ coupling $g'$ then we will have 
$m_{\phi}^2 < 0$, which we will assume in the following analysis.

If we look for an extremum with only  $\vev \phi$  nonzero we find 
\be
\vev{\phi^*\phi} \equiv \half v_{\phi}^2 = \frac{q_{\phi}\xi-m^2_{\phi}}{q^2_{\phi}}
\ee 
so  $\vev{\phi} = O(\sqrt{\xi})$  for large  $\xi$  and 
 $V \approx m_{\phi}^2\xi/q_{\phi}$ . Note that since as indicated 
above we have chosen parameters so that  $m_{\phi}^2 < 0$  
we have $V < 0$ at the minimum. Expanding about the minimum, ie with 
$\phi = (v_{\phi} + H(x))/\sqrt{2}$, where $H$ is the (real) physical 
$U'_1$ Higgs, we find 
\bea
V &=&  \frac{m_{\phi}^2\xi}{q_{\phi}}-\frac{m_{\phi}^4}{2q_{\phi}^2}
+ (m_{\phi}^2 + m_{\phib}^2+\half v_{\phi}^2\lambda_1^2)\phib^*\phib
-\frac{e_i}{q_{\phi}}m_{\phi}^2 \chi^*_i\chi_i\nn
&+&\half v_{\phi}^2\lambda_1^2s^* s  + \half v_{\phi}^2\lambda_2^2 (\nu^c)^*\nu^c\nn
&+& \frac{1}{2}\left(v_{\phi}q_{\phi}H 
-q_{\phi}\phib^*\phib+e_i \chi^*_i\chi_i\right)^2 
\cdots
\label{eq:vquad}
\eea 
For large  $\xi$  (i.e. large  $v_{\phi}$ ) 
all trace of the  $U'_1$  in the effective low energy lagrangian 
disappears, except for 
contributions to the masses of the matter fields 
which are naturally of the same 
order as the AMSB ones. We can see this 
either by treating the heavy $H$-field as non-propagating and 
eliminating it via its equation of motion, or by noting that the 
quartic $(\chi^*\chi)^2$ $D$-term still present in Eq.~\ref{eq:vquad}
is cancelled (at low energies) 
by the $H$-exchange graph using two $H\chi^*\chi$ 
vertices. In the large $\xi$ limit the 
breaking of $U'_1$ preserves supersymmetry; 
thus the $U'_1$ gauge boson, its gaugino, 
$\psi_H$ and $H$ form a massive supermultiplet 
which decouples from the theory. The fact that supersymmetry 
is good at large $\xi$ protects the light $\chi$ fields from 
obtaining masses of $O(\sqrt{\xi})$ from loop corrections.  
Moreover  $v_{\phi}$  via the superpotential gives 
large \sic\ masses to  $\phib, s$  
and also  $\nu^c$ , thus naturally implementing 
the see-saw mechanism. The low energy theory contains just the 
MSSM  fields with the only modification being the FI -type mass contributions 
proportional to  $m^2_{\phi}$ . This is simply another manifestation of the 
non-decoupling of soft mass corrections from $D$-terms\cite{mur}.

We now need only choose the charges  $L,e$  for the 
lepton doublet and singlet so that the contributions to their 
slepton masses are positive; that is, we choose $L,e > 0$ since 
$m_{\phi}^2 < 0$.  It is easy to show that (modulo electroweak 
breaking) this represents the absolute minimum of the potential 
(note that $\lambda_1$ plays a crucial role here in that 
for $\lambda_1 = 0$ the $D$-flat direction 
$\vev\phi = \vev\phib  >> \sqrt{\xi}$
would lead to an potential unbounded from below). 

The model constructed here is similar in spirit to 
those of Harnik et al~\cite{Harnik:2002et}, 
in that the $U'_1$ breaking is at a high scale so that only the
D-term contributions survive in the low energy theory. 
Like \cite{Harnik:2002et} we
assume that the anomaly mediated contribution to  SUSY breaking is
dominant, something that can be justified in the conformal sequestered
scheme of Luty et al~\cite{Luty:2001zv}. The main difference is that we
use a FI term to trigger the $U'_1$ breaking rather than an F-term. Here
we have just assumed the existence of the FI term as one of the terms
allowed by the symmetries of the theory. We will return elsewhere to a
discussion of how such a term may be generated in an underlying theory.

In the next section we will explore the   
region of the  $(e,L)$  
parameter space such 
that electroweak-breaking via the Higgses is obtained as usual.

\section{The sparticle spectrum}

We turn now to the effective low energy theory. 
Evidently in the scenario described in section~\ref{sec:amsb}, 
we have decoupling of the $U'_1$ at low energies 
so that the anomalous dimensions of the fields are as in the MSSM; 
thus for the Higgses and 3rd generation matter fields we have
(at one loop):
\bea
\lf\gamma_{H_1} & = & 3\lambda_b^2+\lambda_{\tau}^2-\frak{3}{2}g_2^2
-\frak{3}{10}g_1^2
,\nn
\lf\gamma_{H_2} & = & 3\lambda_t^2-\frak32g_2^2-\frak{3}{10}g_1^2
,\nn
\lf\gamma_{L} & = & \lambda_{\tau}^2-\frak32g_2^2-\frak{3}{10}g_1^2
,\nn
\lf\gamma_{Q} & = & \lambda_b^2+\lambda_t^2-\frak83g_3^2-\frak32g_2^2
-\frak{1}{30}g_1^2,\nn
\lf\gamma_{t^c} & = & 2\lambda_t^2-\frak83g_3^2-\frak{8}{15}g_1^2,\nn
\lf\gamma_{b^c} & = & 2\lambda_b^2-\frak83g_3^2-\frak{2}{15}g_1^2,\nn
\lf\gamma_{\tau^c} & = & 2\lambda_{\tau}^2-\frak65g_1^2,
\eea
where $\lambda_{t,b,\tau}$ are the third generation Yukawa couplings. 
For the first two generations we use the same expressions 
but without the Yukawa contributions.

The soft scalar masses are given by 
\bea
\mbar^2_Q & = & m^2_Q -\frak{1}{3}L\xi'\quad
\mbar^2_{t^c} = m^2_{t^c} -(\frak{2}{3}L +e)\xi',\nn
\mbar^2_{b^c} & = & m^2_{b^c} +(\frak{4}{3}L+e)\xi',\quad
\mbar^2_L =  m^2_L +L\xi',\nn
\mbar^2_{\tau^c} & = & m^2_{\tau^c} +e\xi',
\quad \mbar^2_{H_{1,2}}   =  m^2_{H_{1,2}} \mp (e+L)\xi',  
\label{eq:smasses}
\eea
where
\be
m_Q^2=\frak{1}{2}m_0^2\mu\frac{d}{d\mu}\gamma_Q 
= \frak{1}{2}m_0^2 \beta_i \frac{\pa}{\pa\lambda_i} \gamma_Q 
\label{eq:squarks}
\ee
(where $\lambda_i$ includes all gauge and Yukawa couplings) and so on, 
and we have written the effective FI parameter as
\be
\xi' = -\frac{m_{\phi}^2}{q_{\phi}}. 
\ee
The 3rd generation $A$-parameters are given by 
\bea
A_t &=-m_0(\gamma_Q+\gamma_{t^c}+\gamma_{H_2}),\nn
A_b &=-m_0(\gamma_Q+\gamma_{b^c}+\gamma_{H_1}), \nn
A_{\tau} &=-m_0(\gamma_L+\gamma_{\tau^c}+\gamma_{H_1})
\label{eq:apams}
\eea
and we set the corresponding first and second generation quantities to zero. 
The gaugino masses are given by 
\be
M_i = m_0 |\frac{\beta_{g_i}}{g_i}|.
\label{eq:inos}
\ee
The scale of the FI contributions is set by the AMSB contribution to the 
$\phi$-mass, and hence is naturally expected to be the same order as the 
other AMSB contributions. In the examples considered below this is indeed the case. 
Clearly these FI contributions depend on two parameters, $L\xi'$ and $e\xi'$. 
For notational simplicity we will 
set $\xi' = 1(\TeV)^2$ from now on. 

We begin by choosing input values for $m_0$, $\tan\beta$, $L$, $e$ 
and $\hbox{sign}\mu$, and then we calculate the appropriate
dimensionless coupling input  values  at the scale $M_Z$ 
by an iterative procedure involving the sparticle spectrum, 
and the loop corrections to $\alpha_{1\cdots 3}$, $m_t$, $m_b$ and $m_{\tau}$, 
as described in Ref.~\cite{bpmz}. We then determine a given sparticle 
pole mass 
by running the dimensionless couplings up to a certain scale chosen 
(by iteration) to be equal to the pole mass itself, 
and then using Eqs.~\ref{eq:squarks}, 
\ref{eq:apams}, \ref{eq:inos} and including full one-loop corrections from
Ref.~\cite{bpmz}, and two-loop corrections 
to the top quark mass~\cite{Bednyakov:2002sf}.
As in Ref.~\cite{jjk}, 
we have compared the effect of using one, two and three-loop anomalous 
dimensions and $\beta$-functions in the calculations. Note that
when doing the three-loop calculation, we 
use in Eq.~\ref{eq:squarks}, for example, the three loop approximation 
for both $\beta_i$ and $\gamma_Q$, thus including some higher order 
effects. 

We will present results for $\mu > 0$ and  $m_0 = 40\TeV$, for which 
value the gluino mass is around $900\GeV$. 

The allowed region in $(e,L)$ space corresponding to an acceptable vacuum 
is shown in Fig.~1. To define the allowed region, 
we have imposed 
$m_{\tautil} > 82\GeV$, $m_{\nutil_{\tau}} > 49\GeV$ and $m_A > 90\GeV$. 
The region is to a good approximation triangular, with one 
side of the triangle corresponding to $m_A$ becoming too light 
(and quickly imaginary just beyond the boundary, with breakdown of 
the electroweak vacuum) and the other two sides to one of the sleptons 
(usually a stau) becoming too
light. (Note that Ref.~\cite{mwells} sets $e=1$ rather than $\xi'=1$, 
which is why the allowed region in their Figure~1 has a different 
shape; the figures are in fact (roughly) equivalent).

\epsfysize= 4in
\centerline{\epsfbox{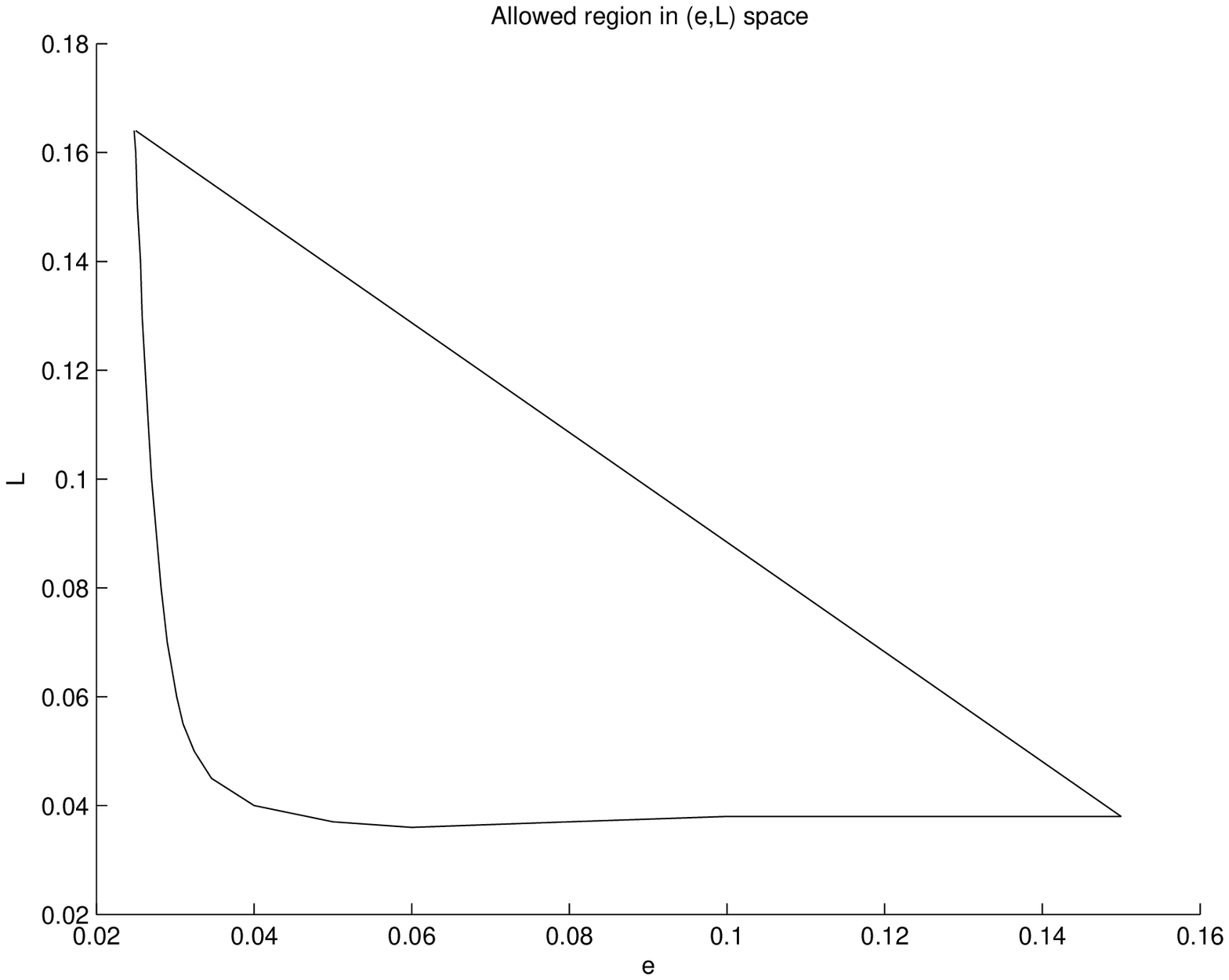}}
\in
{\it \noindent Fig.~1:
The region of $(e,L)$ space corresponding to an acceptable 
electroweak vacuum, for $m_0 = 40\TeV$  
and $\tan\beta = 10$.}
\medskip
\out

Certain features of the spectrum are apparent from Eq.~\ref{eq:smasses}.
Since to avoid tachyonic sleptons 
we must choose $L,e > 0$ we can see that the heaviest squark 
(especially at low $\tan\beta$) is likely to be the mainly right 
handed sbottom. 

As an example of an acceptable spectrum, we give 
the results for $m_0 = 40\TeV, \tan\beta = 10, L = 1/25, e=1/10, 
\hbox{sign}\mu = +$
as derived using the one, two and three loop approximations 
for the anomalous dimensions and $\beta$-functions.  
In Table~\ref{spectrumA} we have used $m_t = 178\GeV$, 
while in Table~\ref{spectrumAb} we have used~\cite{cdf} 
$m_t = 172.7\GeV$. The spectrum is not very much affected by this choice,
the most noticeable alteration being in the mass of the light top squark.
The rest of the results we present will be for  $m_t = 178\GeV$. 
This point in $(e,L)$ space is near the boundary of the allowed region 
(see Fig~1) and is characterised by a light stau. 
The notation in Table~\ref{spectrumA} etc. is fairly standard, 
see for example 
Ref.~\cite{bpmz}; note in particular 
that in all our examples $\ttil_1, \btil_1, \tautil_1$ 
refer to the mainly lefthanded particles. 

\begin{table}
\begin{center}
\begin{tabular}{| c | c | c | c |} \hline
 {\rm mass (GeV)}  &  1{\rm loop}   & 2{\rm loops} &  3{\rm loops}
 \\ \hline 
$ {\tilde g}$  &  914  &  891  & 888  \\ \hline
$ \ttil_1$  &  770 &  761 &  751 \\ \hline
$ \ttil_2$  &  543 &  540 &  529 \\ \hline
$ \util_L $  &  834 &  820 &  809 \\ \hline
$ \util_R $  &  767 &  756 &  744 \\ \hline
$ \btil_1$  &  738  & 728  &  718 \\ \hline   
$\btil_2$  & 928 & 920  & 910  \\ \hline
$ \dtil_L$  &  838  &  824 & 813\\ \hline
$\dtil_R$  &  937  & 929 & 919  \\ \hline
$\tautil_1$  & 124   & 109  &  110  \\ \hline
$\tautil_2$  & 284   & 284  & 284   \\ \hline
$ \etil_L$  &  132  & 118 & 119 \\ \hline
$\etil_R$  &  285 & 285  & 285 \\ \hline
$\nutil_e $  & 104  &  86  & 86  \\ \hline
$\nutil_{\tau} $  & {99}  & {79}  & {80}  \\ \hline
$\chi_1      $  &  105  & 129 & 129  \\ \hline
$\chi_2      $  & 354  &  362 &  361  \\ \hline
$\chi_3      $  &  540 &  563  &  555  \\ \hline
$\chi_4      $  &  552 &  575  & 566  \\ \hline
$\chi^{\pm}_1$  & 106 & 129  &  129  \\ \hline
$\chi^{\pm}_2$   & 549  &  572  &563 \\ \hline
$h      $  & 117  & 117   &  117  \\ \hline  
$H      $  & 315  & 351   &  336  \\ \hline
$A      $  & 314  & 351   & 336  \\ \hline
$H^{\pm}$ & 324 & 360  & 345  \\ \hline
$\chi^{\pm}_1-\chi_1$ (MeV)  &  {230}  & {240} & {240} \\ \hline
\end{tabular}
\caption{\label{spectrumA}
Mass spectrum for $m_t = 178\GeV$, $m_0 = 40\TeV$, $\tan\beta = 10$,
$L =  1/25$,  $e = 1/10$}
\end{center}
\end{table}

\begin{table}
\begin{center}
\begin{tabular}{| c | c | c | c |} \hline
 {\rm mass (GeV)}  &  1{\rm loop}   & 2{\rm loops} &  3{\rm loops}
 \\ \hline 
$ {\tilde g}$  &  914  &  890  & 888  \\ \hline
$ \ttil_1$  &  767 &  758 &  748 \\ \hline
$ \ttil_2$  &  519 &  516 &  505 \\ \hline
$\util_L$  &  835 &  820 &  809 \\ \hline
$ \util_R $  & 767  &  756  &  744 \\ \hline
$ \btil_1$  &  732  & 723  &  713 \\ \hline   
$\btil_2$  & 928 & 920  & 910  \\ \hline
$ \dtil_L$  &  838  &  824 & 813\\ \hline
$\dtil_R$  &  937  & 929 & 919  \\ \hline
$\tautil_1$  & 123   & 108  &  109  \\ \hline
$\tautil_2$  & 284   & 284  & 284   \\ \hline
$ \etil_L$  &  132  & 118 & 119 \\ \hline
$\etil_R$  &  285 & 285  & 285 \\ \hline
$\nutil_e $  & 104  &  85  & 86  \\ \hline
$\nutil_{\tau} $  & 99  & 79  & 80  \\ \hline
$\chi_1      $  &  106  & 129 & 130 \\ \hline
$\chi_2      $  & 355  &  362 &  362  \\ \hline
$\chi_3      $  &  563 &  584  &  576  \\ \hline
$\chi_4      $  &  574 &  595  & 587 \\ \hline
$\chi^{\pm}_1$  & 106 & 130  &  130  \\ \hline
$\chi^{\pm}_2$   & 571  &  592  &584 \\ \hline
$h      $  & 116  & 115   &  115  \\ \hline  
$H      $  & 354  & 385   &  372  \\ \hline
$A      $  & 353  & 384   & 372 \\ \hline
$H^{\pm}$ & 362 & 393  & 381  \\ \hline
$\chi^{\pm}_1-\chi_1$ (MeV)  &  {220}  & {230} & {230} \\ \hline
\end{tabular}
\caption{\label{spectrumAb}
Mass spectrum for $m_t = 172.7\GeV$, $m_0 = 40\TeV$, $\tan\beta = 10$,
$L =  1/25$, $e = 1/10$}
\end{center}
\end{table}

The results exhibit the same feature as found for the Snowmass Benchmark 
(SPS) points in Ref.~\cite{jjk}; that is, the effect of using 3-loop 
$\beta$-functions has a surprisingly large effect on the squark spectrum. 
This effect was most marked in the SPS case when the  
gluino mass was significantly larger than the squark masses, 
which is not the case here; nevertheless, for the light top squark, 
for example,  it is still noticeable. 

A characteristic feature of AMSB distinguishing it from MSUGRA is that
$M_2 < M_1$, where $M_{1,2}$ are the bino and wino masses respectively.
As a result the lightest neutralino (often the LSP) is predominantly 
the neutral wino  and the lighter chargino (often the NLSP) is almost 
degenerate with it. In Table~\ref{spectrumA} we have given all results for masses 
to the nearest $\GeV$; however we have calculated the 
$\chi^{\pm}$, $\chi$ masses using the full one-loop results and 
expect our results for $\chi^{\pm}_1- \chi_1$ to be good to $10\MeV$ as quoted. 
A clear account of the dominant contribution 
to wino mass splitting and the associated 
phenomenology appears in Ref.~\cite{ggw}.  Our splitting 
of around $240\MeV$ 
is consistent with their results.  




Note that in Table~\ref{spectrumA} the $\tau$-sneutrino is the LSP; 
it is interesting 
that the argument of Ref.~\cite{hebb}, which excluded the possibility 
of a sneutrino LSP in the MSUGRA scenario, does not apply here. 
The claim was~\cite{hebb} that with 
MSUGRA boundary conditions, a sneutrino LSP would 
necessarily have a mass less than half the $Z$-mass and so contribute to 
the invisible $Z$ width. Evidently that is not the case here. 
For a recent discussion of sneutrinos as dark matter see~\cite{hmrw}.

An interesting feature of the results is that for low $\tan\beta$ 
we find that the light CP-even Higgs mass, $m_h$, is less than the 
experimental (standard model) lower bound of $114\GeV$. 
Although the generally quoted \sic\ bound is significantly lower, 
we  must take seriously the SM bound here, 
since we find generally that for us $\sin(\beta-\alpha) \sim 1$, so that 
$h$ couples to the $Z$-boson like the SM Higgs. However as 
$\tan\beta$ is increased, $m_h$ increases above this bound.
The allowed range of $\tan\beta$ depends on the choice of $L,e$. 
In Fig~2,3  we plot $m_h$ and the CP-odd Higgs mass $m_A$ against 
$\tan\beta$ for $L = 1/25, e=1/10$. The electroweak vacuum fails 
for $\tan\beta > 25$ in this case. We also plot the lighter stau mass 
(Fig.~4) and the tau sneutrino mass (Fig.~5) 
against $\tan\beta$. We see that acceptable values of 
$m_h$ are obtained for $7 < \tan\beta < 25$, and of the stau mass 
for  $\tan\beta < 19$. 
As can be seen 
from Table~\ref{spectrumA}, $m_h$ is actually 
essentially unchanged by whether we use two or three-loop 
$\beta$-functions; in fact we have used the two-loop $\beta$-functions 
to generate Figs.~2,3. 

In Table~\ref{spectrumB}, we give the results for another 
point in $(e,L)$-space, chosen to be in the centre of the allowed region, 
where this time the lightest neutralino is the LSP.


\begin{table}
\begin{center}
\begin{tabular}{| c | c | c | c |} \hline
 {\rm mass (GeV)}  &  1{\rm loop}   & 2{\rm loops} &  3{\rm loops}
 \\ \hline 
$ {\tilde g}$  &  914  &  890  & 888  \\ \hline
$ \ttil_1$  &  762 & 753 &  743 \\ \hline
$ \ttil_2$  &  554 &  541 &  530 \\ \hline
$\util_L$  &  826&  812 &  801 \\ \hline
$ \util_R $  & 769  &  758  &  746 \\ \hline
$\btil_1$  &  728  & 719  &  709 \\ \hline
$ \btil_2$  & 940 & 932 & 923  \\ \hline
$\dtil_L$  &  830  &  816 & 805  \\ \hline
$ \dtil_R$  &  949  & 941 & 932  \\ \hline
$\tautil_1$  & 212   & 208  &  208  \\ \hline
$\tautil_2$  & 250   & 247  & 247    \\ \hline
$\etil_L$  &  228  & 228 & 228 \\ \hline
$ \etil_R$  &  241 & 234  & 235 \\ \hline
$\nutil_e $  & 227  &  220  &  220  \\ \hline
$\nutil_{\tau} $  & 225  & 218  & 218  \\ \hline
$\chi_1      $  &  106  &  130& 130  \\ \hline
$\chi_2      $  & 353  &  361 &  361  \\ \hline
$\chi_3      $  &  530 &  554  &  545  \\ \hline
$\chi_4      $  &  543 &  566  & 557 \\ \hline
$\chi^{\pm}_1$  & 106 & 130 &  130  \\ \hline
$\chi^{\pm}_2$   & 539 & 562  & 553 \\ \hline
$h      $  & 117  & 117   &  117  \\ \hline  
$H      $  & 277  & 319   &  303  \\ \hline
$A      $  & 277  & 319   & 302  \\ \hline
$H^{\pm}$ & 288 & 329 & 313  \\ \hline
$\chi^{\pm}_1-\chi_1$ (MeV)  & 240   & 250 & 250  \\ \hline
\end{tabular}
\caption{\label{spectrumB}
Mass spectrum for $m_0 = 40\TeV$, $\tan\beta = 10$,
$L =  0.08$, $e = 0.07$
}
\end{center}
\end{table}

Finally in Table~\ref{spectrumC} 
we give results for $(e,L) = (0.05,0.05)$, a point again 
near the boundary in $(e,L)$ space, with 
light sleptons and heavy squarks, and also a 
large charged Higgs mass of over $400\GeV$. This point is interesting because 
of the fact that previous authors have noted 
that the fact that $M_3$ and $M_2$ have opposite signs 
disfavours at first sight
a \sic\ explanation of the well-known discrepancy
between theory and experiment for  the anomalous magnetic moment
of the muon, $a_{\mu}$. This is
because if sign ($\mu M_2$) is chosen so as to create
a positive $a_{\mu}^{\rm{SUSY}}$ then sign ($\mu M_3$) leads to
constructive interference between various \sic\ contributions to
$B(b \to s\gamma)$, and consequent restrictions on the allowed
parameter space. However, with light sleptons (to generate 
a contribution to $a_{\mu}$) and heavy squarks and charged Higgs 
(to suppress contributions to $B(b \to s\gamma)$) 
this conclusion can be evaded (as was already argued in Ref.~\cite{jjyt}).

There has been a considerable amount of work in recent years on  two
loop corrections to $m_h$~\cite{sven}-\cite{martin}. For some regions of
the MSSM parameter space these can be substantial; therefore since we
have presented  predictions of around $115-118\GeV$ we have to worry
about them since  they generally reduce $m_h$. Using the useful web
resource from  Ref.~\cite{sven}, we obtain, for the input parameters of 
Table~\ref{spectrumC}, the result $m_h = 116.2\pm 1.4\GeV$,  in
excellent agreement with our results,  which suggests that the two-loop
corrections are in fact  not very large in our scenario.  Other points
in $(e,L)$ space give similar results.  Thus for $m_0 = 40\TeV$ we 
predict that $m_h$ is less than about $118.4\GeV$ (see Fig.~2). If we 
increase $m_0$ then this bound does increase somewhat (to around $125\GeV$
at $m_0 =100\TeV$, for example) but at the price of considerable 
electroweak fine-tuning.

\begin{table}
\begin{center}
\begin{tabular}{| c | c | c | c |} \hline
 {\rm mass (GeV)}  &  1{\rm loop}   & 2{\rm loops} &  3{\rm loops}
 \\ \hline 
$ {\tilde g}$  &  914  &  890  & 888  \\ \hline
$ \ttil_1$  &  772 & 763 &  753 \\ \hline
$ \ttil_2$  &  579 &  576 &  566 \\ \hline
$ \util_L $  & 832&  818 &  807 \\ \hline
$\util_R$  &  795  &  785  &  773 \\ \hline 
$\btil_1$  & 737 & 727 & 718  \\ \hline
$ \btil_2$  &  909  & 900  & 890 \\ \hline
$\dtil_L$  &  836  & 822 & 811 \\ \hline
$ \dtil_R$  &  917  &  909 & 899  \\ \hline
$\tautil_1$  & 140   & 130  & 131    \\ \hline
$\tautil_2$  & 194   & 191  & 191   \\ \hline
$\etil_L$  &  168 & 158  & 158 \\ \hline
$ \etil_R$  &  178  & 177 & 177 \\ \hline
$\nutil_e $  & 148  &  137  &  137  \\ \hline
$\nutil_{\tau} $  & 144  & 133  & 133  \\ \hline
$\chi_1      $  &  106  &  130 & 130  \\ \hline
$\chi_2      $  & 355  &  362 &  362  \\ \hline
$\chi_3      $  &  577 &  599  &  590  \\ \hline
$\chi_4      $  &  587 &  609  & 601 \\ \hline
$\chi^{\pm}_1$  & 106 & 130 &  130 \\ \hline
$\chi^{\pm}_2$   & 585 & 606  & 598 \\ \hline
$h      $  & 117  & 117   &  117  \\ \hline  
$H      $  & 429 & 455   &  444  \\ \hline
$A      $  & 428  & 454   & 443  \\ \hline
$H^{\pm}$ & 436 & 462 & 451  \\ \hline
$\chi^{\pm}_1-\chi_1$ (MeV)  & 220   & 230 & 230  \\ \hline
\end{tabular}
\caption{\label{spectrumC}
Mass spectrum for $m_0 = 40\TeV$, $\tan\beta = 10$,
$L =  0.05$, $e = 0.05$
}
\end{center}
\end{table}

\epsfysize= 4in
\centerline{\epsfbox{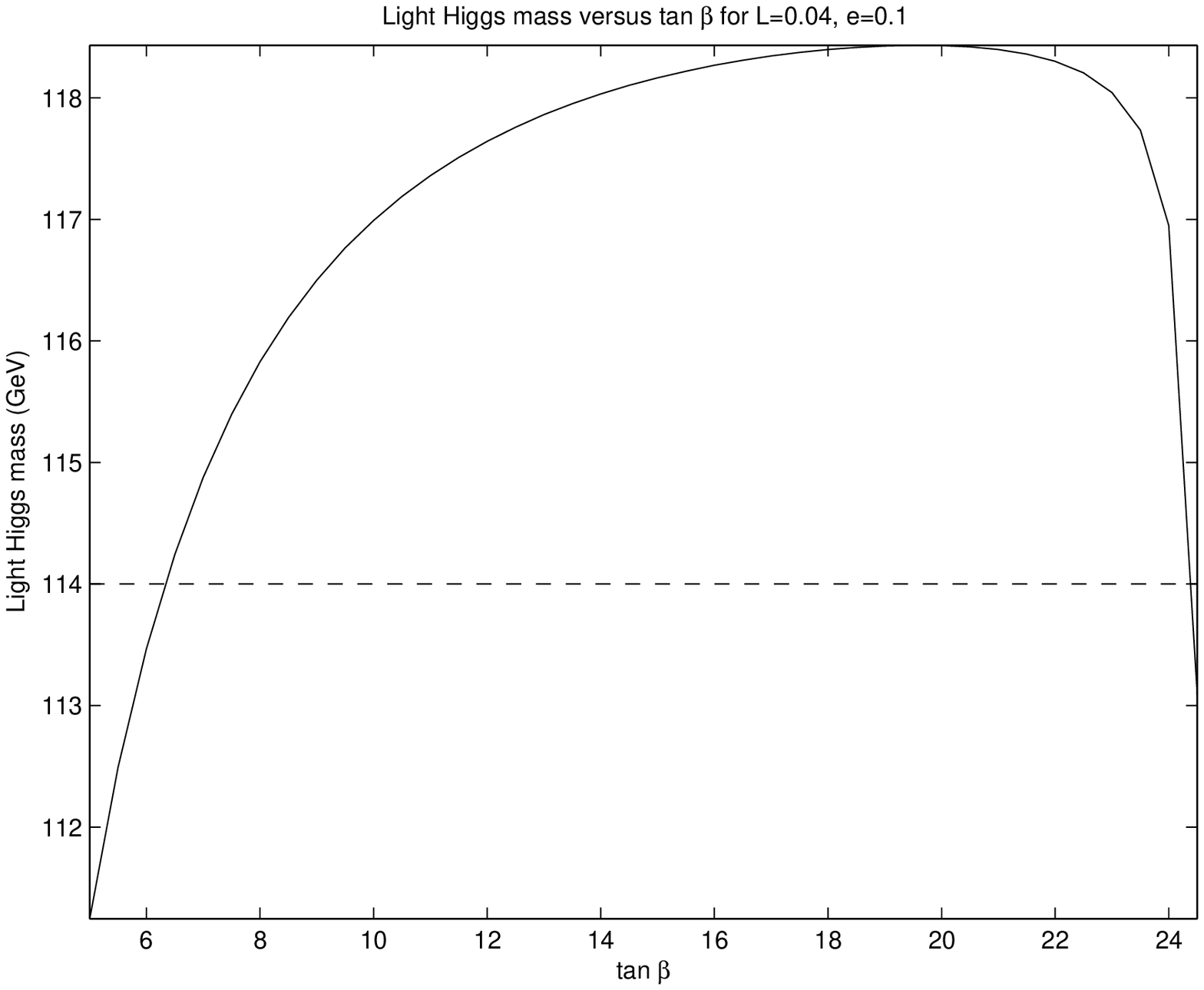}}
{\it \noindent Fig.~2:
The light CP-even Higgs mass $m_h$ as a function of $\tan\beta$, 
for $L=1/25, e=1/10$. 
The dotted line is the SM lower limit $(114\GeV)$.
}\medskip

\epsfysize= 4in
\centerline{\epsfbox{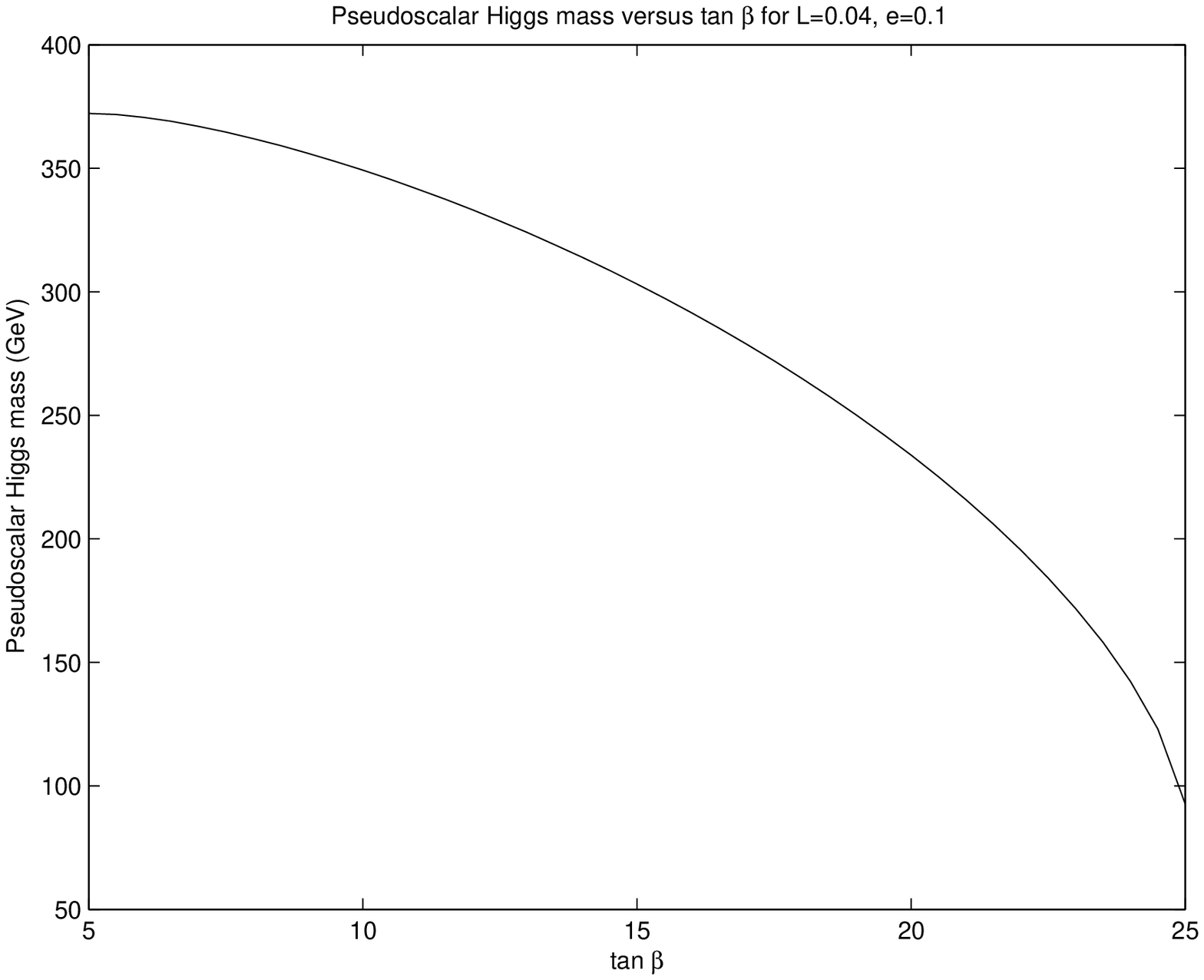}}
\label{fig:higgs2}
{\it \noindent Fig.~3:
The CP-odd Higgs mass $m_A$ as a function of $\tan\beta$, 
for $L=1/25, e=1/10$.
}
\medskip

\epsfysize= 4in
\centerline{\epsfbox{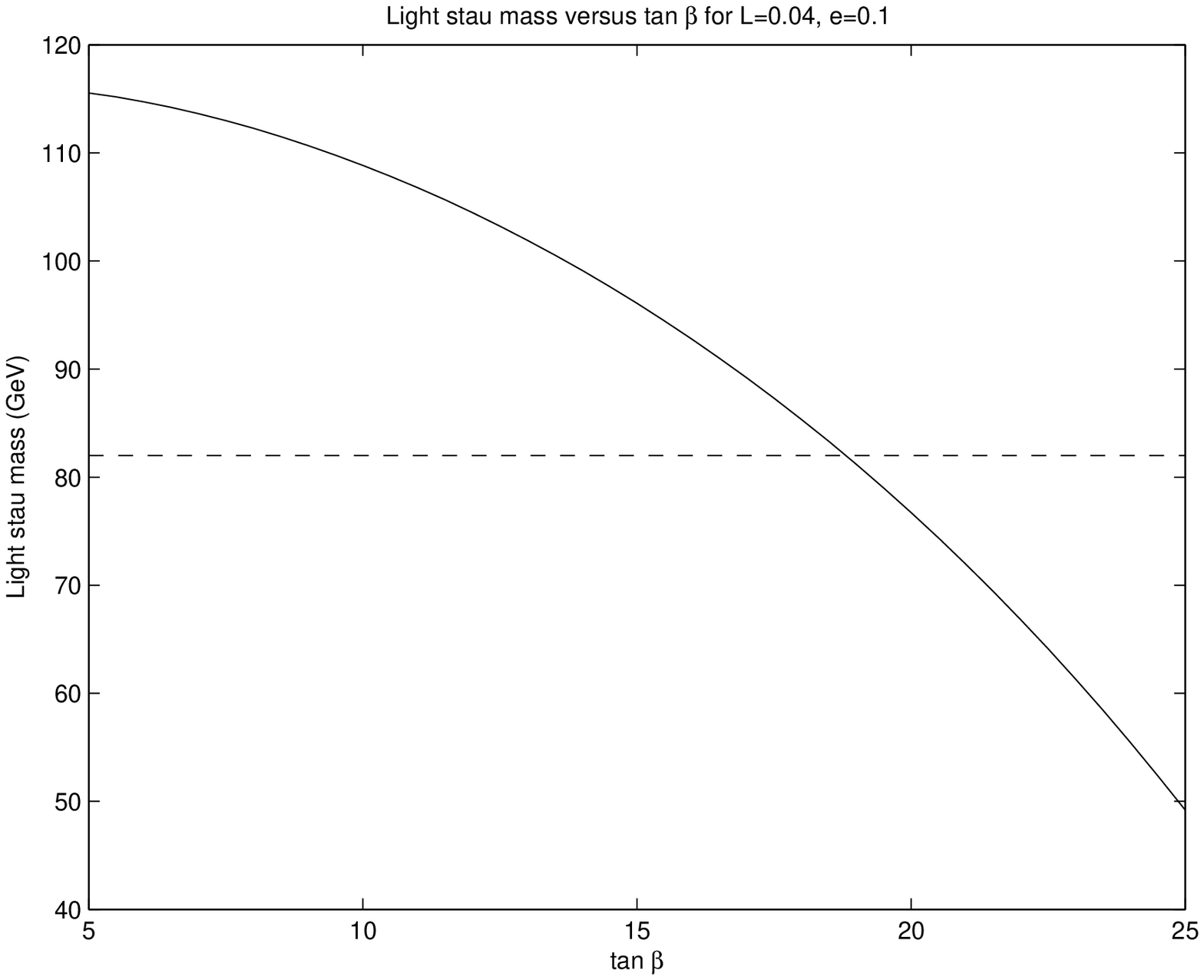}}
{\it \noindent Fig.~4:
The light stau mass as a function of $\tan\beta$,
for $L=1/25, e=1/10$.
The dotted line is the lower limit $(82\GeV)$.
}\medskip

\epsfysize= 4in
\centerline{\epsfbox{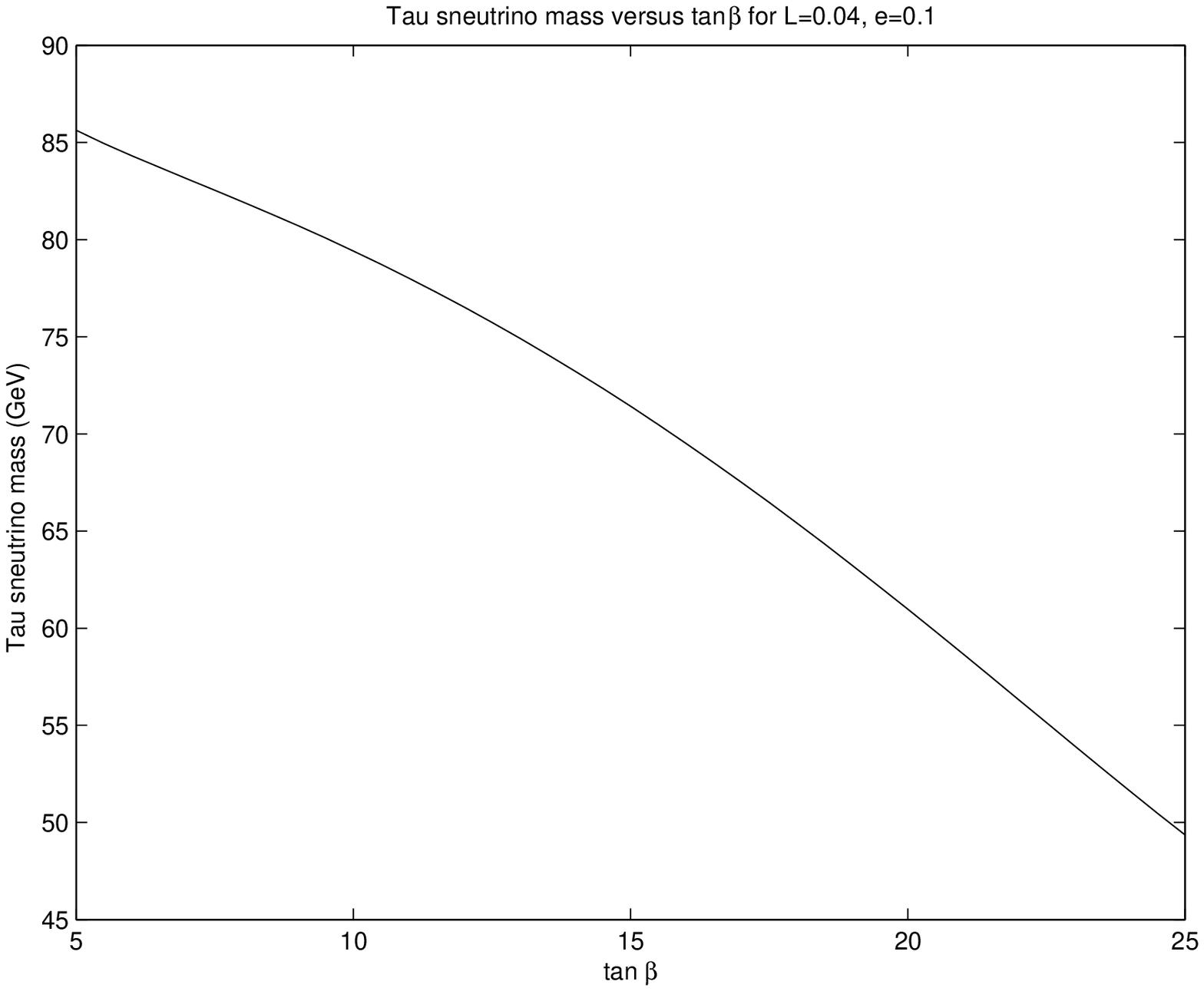}}
{\it \noindent Fig.~5:
The $\tau$-sneutrino mass as a function of $\tan\beta$,
for $L=1/25, e=1/10$.
}\medskip

\section{Mass sum rules}

By taking appropriate linear combinations of masses it is straightforward 
to derive a set of interesting sum rules~\cite{jja}.

In the following equations, if we substitute the tree values for the various 
masses on the left hand side,  the $(e,L)$ dependent terms and the 
electroweak breaking contributions to the masses  cancel. We have calculated the 
numerical coefficients on the right hand side from Table~2, using the 
two-loop sparticle mass predictions; it is easy to then check that to the indicated 
accuracy the same equations hold for the results in Tables~4, 5.
Thus these sum rules are to an excellent approximation independent 
of $(e,L)$, and also in fact of $m_0$; 
the numerical coefficients are  slowly varying functions of $\tan\beta$ and the input 
top pole mass.


\bea
m_{\ttil_1}^2+m_{\ttil_2}^2+m_{\btil_1}^2+m_{\btil_2}^2
- 2m_t^2 & = &
2.76\left(m_{\gtil}\right)^2  \nn
m_{\tautil_1}^2+m_{\tautil_2}^2+m_{\ttil_1}^2+m_{\ttil_2}^2
- 2m_t^2 & = & 1.14\left(m_{\gtil}\right)^2.
\label{eq:suma}
\eea

\bea
m_{\etil_L}^2+2m_{\util_L}^2+m_{\dtil_L}^2 & = &
2.60\left(m_{\gtil}\right)^2,\nn
m_{\util_R}^2+m_{\dtil_R}^2+m_{\util_L}^2+m_{\dtil_L}^2 & = &
3.51\left(m_{\gtil}\right)^2,\nn
m_{\util_L}^2+m_{\dtil_L}^2-m_{\util_R}^2-m_{\etil_R}^2 & = &
0.88\left(m_{\gtil}\right)^2.
\label{eq:sumb}
\eea

\bea
m_A^2 - 2\sec 2\beta\left (m_{\etil_L}^2 +m_{\etil_R}^2\right) & = &
0.40\left(m_{\gtil}\right)^2,\nn
m_A^2 - 2\sec 2\beta\left (m_{\tautil_1}^2 + m_{\tautil_2}^2 
-2m_{\tau}^2\right) & = &
0.39\left(m_{\gtil}\right)^2.
\label{eq:sumc}
\eea

The existence of these sum rules will be a useful distinguishing feature 
of the AMSB scenario. 

\section{Conclusions}

Despite remarkable advances in the understanding of string theory, 
a coherent high energy theory spawning the MSSM as an effective low 
energy theory remains elusive. This has led to exploration of such 
outr\'e possibilities as little higgs models and split supersymmetry.
Remaining within the conservative world of low energy \sy, the AMSB 
scenario is an attractive alternative to (and easily 
distinguished from) MSUGRA. We have shown how a $U'_1$ gauge symmetry 
broken at high energies can lead in a natural way to the 
FI-solution to the tachyonic slepton problem in the context 
of anomaly mediation. The result is a sparticle spectrum 
described by the parameter set $m_0,e,L, \tan\beta, \hbox{sign}(\mu)$;
and it is only for a comparatively restricted set of $(e,L)$ 
that an acceptable spectrum is obtained. 
Moreover we have presented a set of sum rules which are independent 
of $m_0$, $L$ and $e$.
At the very least, the scenario we describe 
has the merit of being immediately testable should sparticles 
be discovered in experiments at the LHC.

\section*{\large Acknowledgements}

DRTJ was supported by a PPARC Senior Fellowship, and a CERN Research
Associateship, and both he and GGR were 
visiting CERN while part of this work was done.
DRTJ thanks Howie Haber for correspondence, Sabine Kraml for 
conversations and Sven Heinemeyer for 
a patient introduction to the simplicities of {\it FeynHiggs\/}.
This work was partially supported by the EC 6th Framework Programme
MRTN-CT-2004-503369.

\end{document}